\documentclass{ws-procs9x6}

\begin{document}

\title{Combined LOPES and KASCADE-Grande Data Analysis%
%\footnote{\em haungs@ik.fzk.de}
}

%\author{ANDREAS HAUNGS}
%\address{Institut\ f\"ur Kernphysik, Forschungszentrum Karlsruhe, Germany 
%{\em haungs@ik.fzk.de}}

\author{   
  A.~HAUNGS$^{a}$,
  W.D.~APEL$^{a}$,
  F.~BADEA$^{a}$,
  L.~B\"AHREN$^{b}$,
  K.~BEKK$^{a}$, 
  A.~BERCUCI$^{c}$,
  M.~BERTAINA$^{d}$,
  P.L.~BIERMANN$^{e}$,
  J.~BL\"UMER$^{a,f}$,
  H.~BOZDOG$^{a}$,
  I.M.~BRANCUS$^{c}$,
  M.~BR\"UGGEMANN$^{g}$,
  P.~BUCHHOLZ$^{g}$,
  S.~BUITINK$^{h}$,
  H.~BUTCHER$^{b}$,
  A.~CHIAVASSA$^{d}$,
  K.~DAUMILLER$^{a}$,
  A.G.~DE~BRUYN$^{b}$,
  C.M.~DE~VOS$^{b}$,
  F.~DI~PIERRO$^{d}$,
  P.~DOLL$^{a}$, 
  R.~ENGEL$^{a}$,
  H.~FALCKE$^{b,e,h}$,
  H.~GEMMEKE$^{i}$,
  P.L.~GHIA$^{j}$,
  R.~GLASSTETTER$^{k}$, 
  C.~GRUPEN$^{g}$,
  D.~HECK$^{a}$, 
  J.R.~H\"ORANDEL$^{f}$,
  A.~HORNEFFER$^{h,e}$,
  T.~HUEGE$^{a,e}$,
  K.-H.~KAMPERT$^{k}$,
  G.W.~KANT$^{b}$,
  U.~KLEIN$^{l}$,
  Y.~KOLOTAEV$^{g}$,
  Y.~KOOPMAN$^{b}$,
  O.~KR\"OMER$^{i}$,
  J.~KUIJPERS$^{h}$,
  S.~LAFEBRE$^{h}$,
  G.~MAIER$^{a}$,
  H.J.~MATHES$^{a}$, 
  H.J.~MAYER$^{a}$, 
  J.~MILKE$^{a}$, 
  B.~MITRICA$^{c}$,
  C.~MORELLO$^{j}$,
  G.~NAVARRA$^{d}$,
  S.~NEHLS$^{a}$, 
  A.~NIGL$^{h}$,
  R.~OBENLAND$^{a}$,
  J.~OEHLSCHL\"AGER$^{a}$, 
  S.~OSTAPCHENKO$^{a}$, 
  S.~OVER$^{g}$,
  H.J.~PEPPING$^{b}$, 
  M.~PETCU$^{c}$,
  J.~PETROVIC$^{h}$, 
  T.~PIEROG$^{a}$, 
  S.~PLEWNIA$^{a}$, 
  H.~REBEL$^{a}$, 
  A.~RISSE$^{m}$, 
  M.~ROTH$^{f}$, 
  H.~SCHIELER$^{a}$, 
  G.~SCHOONDERBEEK$^{b}$,
  O.~SIMA$^{c}$, 
  M.~ST\"UMPERT$^{f}$, 
  G.~TOMA$^{c}$, 
  G.C.~TRINCHERO$^{j}$,
  H.~ULRICH$^{a}$,
%  S.~VALCHIEROTTI$^{d}$,
  J.~VAN~BUREN$^{a}$,
  W.~VAN~CAPELLEN$^{b}$,
  W.~WALKOWIAK$^{g}$,
  A.~WEINDL$^{a}$,
  S.~WIJNHOLDS$^{b}$,
  J.~WOCHELE$^{a}$, 
  J.~ZABIEROWSKI$^{m}$,
  J.A.~ZENSUS$^{e}$,
  D.~ZIMMERMANN$^{g}$
     \\ LOPES COLLABORATION
}
\address{
$^{A}$ Institut\ f\"ur Kernphysik, Forschungszentrum Karlsruhe, Germany\\
$^{B}$ ASTRON Dwingeloo, The Netherlands\\
$^{C}$ NIPNE Bucharest, Romania\\
$^{D}$ Dpt di Fisica Generale dell'Universit{\`a} Torino, Italy\\
$^{E}$ Max-Planck-Institut f\"ur Radioastronomie, Bonn, Germany\\
$^{F}$ Institut f\"ur Experimentelle Kernphysik, Uni Karlsruhe, Germany,\\
$^{G}$ Fachbereich Physik, Universit\"at Siegen, Germany \\
$^{H}$ Dpt of Astrophysics, Radboud Uni Nijmegen, The Netherlands\\
$^{I}$ IPE, Forschungszentrum Karlsruhe, Germany\\
$^{J}$ Ist di Fisica dello Spazio Interplanetario INAF, Torino, Italy\\
$^{K}$ Fachbereich Physik, Uni Wuppertal, Germany \\
$^{L}$ Radioastronomisches Institut der Uni Bonn, Germany\\
$^{M}$ Soltan Institute for Nuclear Studies, Lodz, Poland
}  

\maketitle

\abstracts{
First analyses of coincident data of the LOPES (LOfar PrototypE Station) 
radio antennas with the particle air shower experiment 
KASCADE-Grande show basic correlations in the observed shower parameters, 
like the strength of the radio signal and the particle number, or comparing 
the estimated shower directions. In addition, an improvement of the 
experimental resolution of the shower parameters reconstructed by 
KASCADE-Grande can be obtained by including the data of the radio antennas. 
This important feature will be shown in this article explicitely by an example
event.}

%\section{LOPES and KASCADE-Grande}
The KASCADE\cite{Anton03} (KArlsruhe Shower Core and Array DEtector) 
experiment measures showers in a primary energy range from \mbox{100 TeV} 
to \mbox{80 PeV} and provides multi-parameter measurements on a large 
number of observables concerning electrons, muons at 4 different energy 
thresholds, and hadrons.
The main detector components of KASCADE are a field array, the so called 
central detector and a muon tracking detector. The field array consists of 
252 detector stations with shielded as well as unshielded scintillation 
detectors for measuring the electromagnetic and the muonic shower component. \\
KASCADE-Grande\cite{Navar03} is the extension of the multi-detector 
setup KASCADE to cover a primary cosmic ray energy range from 
\mbox{100 TeV} to \mbox{1 EeV}. 
Grande is an array of \mbox{700 x 700 m$^2$} equipped with 37 plastic 
scintillator stations sensitive to measure energy deposits and arrival times 
of air shower particles. \\
At present, LOPES\cite{Falck05} operates 30 dipole radio antennas 
(LOPES-30) positioned inside or nearby KASCADE. 
For the present analysis only data of LOPES-10 
(10 antennas in operation), which was running for 5 months, is used. 
The antennas operate in the frequency range of 
\mbox{40-80 MHz}. The radio data is collected when a "large event" trigger is 
received from KASCADE which translates to primary energies above 
\mbox{$10^{16}$ eV}.  \\

%\section{Data Analysis}
The LOPES-10 data set is subject of various analyses using different 
selections: With an event sample obtained by hard cuts the proof of principle 
to detect air showers in the radio frequency range was given\cite{Falck05}. 
With events fallen inside KASCADE the basic correlations with shower 
parameters are shown\cite{Horn05}.
Further interesting features are investigated with a sample of very inclined 
showers\cite{jelena} and with a sample of events measured during 
thunderstorms\cite{steijn}.  \\
Here we report results from an analysis performed by correlating the radio 
signals measured by LOPES-10 with EAS events reconstructed by KASCADE-Grande 
with remote cores included\cite{badea}.
Grande is taking data in coincidence with KASCADE and LOPES
and enables to reconstruct showers with primary energies up to \mbox{10$^{18}$ 
eV} and with distances between shower core and the LOPES-10
antennas up to \mbox{700 m}. 
The Grande reconstruction accuracy of shower core position and direction is in 
the order of \mbox{4 m} (\mbox{13 m}) and $0.18^{\circ}$ ($0.32^{\circ}$) with 
$68$\% ($95$\%) confidence level for simulated proton and iron showers at 
\mbox{100 PeV} primary energy and $22^{\circ}$ zenith 
angle\cite{Glass03}. \\

A crucial element of the detection method is the digital beamforming which 
allows to place a narrow antenna beam in the direction of the cosmic ray event.
This is possible because the phase information of the radio waves is preserved 
by the digital receiver and the cosmic ray produces a coherent pulse. 
This method is also very effective in suppressing interference from the 
particle detectors which radiate incoherently. 
The procedure of time shifting of the radio signals in the antennas is 
relatively safe when based on the values provided by the reconstruction for 
shower core and shower axis using the data of the original KASCADE field array. 
Due to the high granularity of the detector stations the accuracy of core and 
direction reconstruction is high enough to obtain a good coherence of the
radio signals. 
%Of course, this is valid only for showers with their cores inside 
%KASCADE and with not too energetic showers which may lead to saturation of 
%the detectors. \\
The shower reconstruction using information from the Grande array is required 
for shower cores outside KASCADE. The Grande stations, \mbox{10 m$^2$} of plastic 
scintillator detectors each, are spaced at \mbox{$\simeq$130 m} and cannot assure 
an accuracy comparable with the original KASCADE array. So, a so-called 'optimised' 
beamforming is performed, which searches for a maximum coherence by varying the 
core and the direction around the values provided by the Grande reconstruction. 

\begin{figure}[ht]
\centerline{\epsfxsize=5.5cm\epsfbox{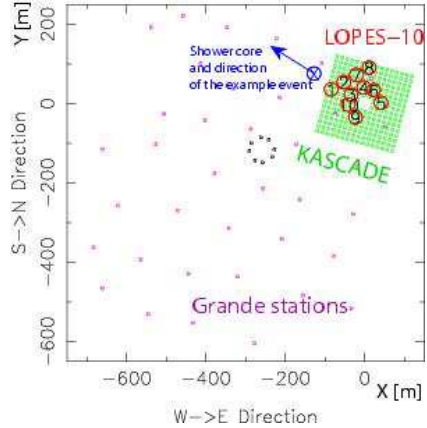}}
\caption{Sketch of the LOPES~10 layout inside KASCADE-Grande. 
Core position and shower direction of the candidate event 
444 is also shown. \label{event444}}
\end{figure}

\begin{figure}[ht]
\centerline{\epsfxsize=8.5cm   
\epsfbox{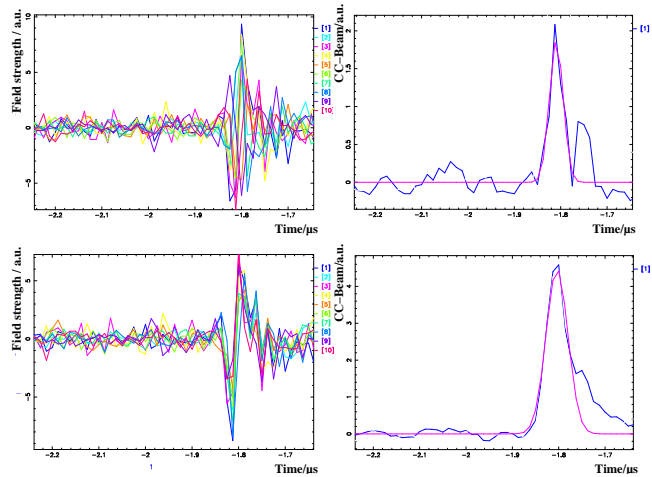}}
\caption{Radio signals in the 10 antennas and the resulting radio signal 
(so-called cross-correlation beam) with a gaussian fit to the signal
after first beamforming (upper panels) and after {\it optimised}
beamforming (lower panels). \label{resolution}}
\end{figure}

%\section{Data selection}
In five months of LOPES-10 data taking a sample of 862 candidate events were
selected. Selection criteria were 
i) coincident measurement of the event by LOPES~10, 
KASCADE field array which have triggered LOPES, and Grande array;
ii) zenith angle of the shower less than $50^\circ$; 
iii) a geometrical cut that the core position lies inside the Grande array 
($0.358\,$km$^2$); 
iv) to reduce the data sample additionally an energy and distance cut is 
applied which is motivated by Allan's fomula\cite{Allan71}:
\begin{small}
$\epsilon_\nu=20 \cdot \left(\frac{E}{10^{17}eV}\right) \cdot \sin{\alpha}
\cdot \cos{\theta} \cdot
\exp\left(-\frac{R}{R_0(\nu,\theta)}\right)\,\,
\left[\frac{\mu V}{m\cdot MHz}\right]$.
\end{small}
The formula describes the pulse amplitude per unit bandwidth ($\epsilon_\nu$) of  
the radio signal induced by an EAS.
Here $E$ is the primary energy, $\alpha$ the angle to the geomagnetic field, 
$\theta$ the zenith angle, $R$ the distance to the shower axis and $R_0$ the 
scaling radius \mbox{$R_0=110$ m} at \mbox{55 MHz}; the exponential radial factor 
plays a significant role for remote showers.
The cut has been considered as
\begin{small}
$ \lg\left(\frac{E}{eV}\right) > \lg\left(\frac{E_0}{eV}\right)+0.4343\cdot 
\frac{R}{R_0}\,\,\,\,\,\,OR\,\,\,\,\,\,\lg\left(\frac{E}{eV}\right)>17.5$
\end{small}
with \mbox{$E_0=10^{16.5}$ eV} and \mbox{$R_0=160$ m}, i.e. weaker than 
Allan's scaling with radius. \\
%\section{Event example}
Fig.~\ref{event444} shows the layout of LOPES-10 in KASCADE-Grande including the
core position and direction of an event (number 444 of 862) with a 
clear radio signal. Fig.~\ref{resolution}
shows the radio signal of the event before and after the optimized beamforming.
An noticeable increase is seen in the final beam signal after the 
optimised beamforming. Table~\ref{SCA} contains the values (azimuthal
angle $\phi$, zenith angle $\theta$, $X_{core}$ and $Y_{core}$) reconstructed by
Grande and the corresponding values obtained after maximizing the radio
coherence for this event;
the small shifts assure an almost perfect coherence of the radio signal (lower
left panel in Fig.~\ref{resolution}).  

\begin{table}[h]
\tbl{Shower parameters from the Grande reconstruction and after maximized
radio coherence for an example event (geomagnetic angle 
$\alpha=57.9^\circ$, lg(E/eV)$=17.73$).}
{ \begin{small}
\begin{tabular}{|c|c|c|} \hline
parameter  &Grande reconstruction& maximized radio coherence\\
                             \hline 
$\phi$      &  302.2$^\circ$  &299.3$^\circ$\\
$\theta$    &  41.0$^\circ$   &39.9$^\circ$\\
$X_{core}$/m  & -142.8  &-139.2\\
$Y_{core}$/m  &   40.3  &   51.6\\
curvature/m  &   3250  &   4250\\
                             \hline		
\end{tabular}\label{SCA} \end{small}}
\vspace*{-10pt}
\end{table}

The maximization of the radio coherence (optimised beamforming) will be 
performed to all the candidate events. 
This will not only increase the efficiency ($\approx 30$\% without optimized 
beamforming) for finding radio signals of air showers, but also improve the 
reconstruction quality on the primary energy of the cosmic rays if the 
information is included in the shower reconstruction of Grande.


\begin{thebibliography}{0}

\bibitem{Anton03}
T. Antoni et al. - KASCADE collab., {\it Nucl. Instr. \& Meth. A} 
{\bf 513}, 429 (2003).

\bibitem{Navar03}
G. Navarra et al. - KASCADE-Grande collab., {\it Nucl. Instr. \& Meth. A} 
{\bf 518}, 207 (2004).

\bibitem{Falck05} 
H. Falcke et al. - LOPES collab., {\it Nature} {\bf 435}, 313 (2005).

\bibitem{Horn05} 
A. Horneffer et al. - LOPES collab., {\it Detection of radio pulses from extensive air showers}, 
Proc. of 29$^{th}$ ICRC, Pune, India (2005).

\bibitem{jelena}
J. Petrovic et al. - LOPES collab., {\it Radio emission of highly inclined cosmic ray air showers measured with LOPES}, 
Proc. of 29$^{th}$ ICRC, Pune, India (2005). 

\bibitem{steijn}
S. Buitink et al. - LOPES collab., {\it Electric field influence on the radio emission of air showers}, 
Proc. of 29$^{th}$ ICRC, Pune, India (2005); 

\bibitem{badea}
A.F. Badea et al. LOPES collab., {\it Remote event analyses of LOPES-10}, 
Proc. of 29$^{th}$ ICRC, Pune, India (2005); \\
A.F. Badea et al. LOPES collab., {\it First determination of the reconstruction 
resolution of an EAS radio detector}, 
Proc. of 29$^{th}$ ICRC, Pune, India (2005); 

\bibitem{Glass03} 
R. Glasstetter et al. - KASCADE-Grande collab., Proc. of 28$^{th}$ ICRC, Tsukuba, Japan, 781 (2003).

\bibitem{Allan71} 
H.R.~Allan, {\it Prog. in Element. Part. and Cos. Ray Phys.}, {\bf 10}, 171 (1971).

\end{thebibliography}
\end{document}